%% file: main.tex
\documentclass[a4paper]{article}

\usepackage{INTERSPEECH2021}

\usepackage{soul,color}
\usepackage{subcaption}
\usepackage{mathtools}
\usepackage{hyperref}
\usepackage[super]{nth}
\newcommand{\norm}[1]{\left\lVert#1\right\rVert}

\title{Multi-Scale Spectrogram Modelling for Neural Text-to-Speech}
\name{Ammar Abbas, Bajibabu Bollepalli, Alexis Moinet, Arnaud Joly, Penny Karanasou, Peter Makarov, Simon Slangens, Sri Karlapati, Thomas Drugman}

\address{
  Alexa AI, Cambridge, United Kingdom}
\email{\{syeabbs, bajibabb, drugman\}@amazon.com}

\begin{document}

  \maketitle
  \begin{abstract}
    We propose a novel Multi-Scale Spectrogram (MSS) modelling approach to synthesise speech with an improved coarse and fine-grained prosody. We present a generic multi-scale spectrogram prediction mechanism where the system first predicts coarser scale mel-spectrograms that capture the suprasegmental information in speech, and later uses these coarser scale mel-spectrograms to predict finer scale mel-spectrograms capturing fine-grained prosody.
    We present details for two specific versions of MSS called \emph{Word-level MSS} and \emph{Sentence-level MSS} where the scales in our system are motivated by the linguistic units. The Word-level MSS models word, phoneme, and frame-level spectrograms while Sentence-level MSS models sentence-level spectrogram in addition. 
    Subjective evaluations show that Word-level MSS performs statistically significantly better compared to the baseline on two voices.
  \end{abstract}
  \noindent\textbf{Index Terms}: neural text-to-speech, multi-scale spectrogram, word-level, sentence-level

  \input{intro}
  \input{baseline}

  \input{mss}

  \input{experiments}
  \input{conclusions}

  \bibliographystyle{IEEEtran}
  \bibliography{references}

\end{document}

%% file: intro.tex
\section{Introduction}
\label{section:introduction}

Over the last few years, the progress in Text-To-Speech (TTS) technology has been astounding.
Specifically, neural models such as Wavenet \cite{oord2016wavenet} and Tacotron \cite{wang2017tacotron} have revamped all the components of a modern TTS system \cite{zen2009statistical}. 
Due to this, the Neural Text-To-Speech (NTTS) has become a standard paradigm where a neural sequence-to-sequence (seq2seq) model is employed to map the input text into acoustic features, and a neural vocoder model is employed to convert the acoustic features into a corresponding waveform.
These NTTS systems are capable of generating high-quality speech that is often indistinguishable from human speech~\cite{shen2018natural}.

However, NTTS systems still struggle to produce speech with  appropriate prosody compared to human speech~\cite{ning2019review}. 
The perceived prosody in the synthesized speech may sound inappropriate given the textual context, and includes problems such as wrong type of intonation, pausing, or emphasis. The lack of appropriate prosody stems from multiple reasons ranging from the model design to the way data is processed.
Although the prosody is a suprasegmental phenomenon, the existing NTTS systems are designed in a way that they take fine textual representations such as phonemes as an input and predict finer-level acoustic representations such as mel-spectrograms as an output. 
Thus, the speech produced by the NTTS system can sound flat and require more cognitive effort to process it~\cite{curtin2017cognitive}.

Numerous studies have been proposed in the literature to address the aforementioned issues \cite{skerry2018towards,karlapati2020prosodic,hodari2020camp,sun2020generating}.
Most of these studies focus on modelling a latent representation space of prosody using a separate encoder called reference encoder~\cite{skerry2018towards}.
The reference encoder guides the prosody of the output speech signal to generate expressive speech. 
The reference encoder can be designed in a variational \cite{akuzawa2018expressive} or non-variational \cite{skerry2018towards} style.
Moreover, the latent embedding vectors encoded by reference encoder can be represented either at a coarser level e.g. sentence~\cite{karlapati2020prosodic} or at a finer level e.g. word or phonemes~\cite{hodari2020camp,sun2020generating}.

Along with these latent representation based methods, another set of studies focus on modelling prosody in a hierarchical manner along with a reference encoder \cite{hsu2018hierarchical,kenter2019chive,chien2020hierarchical}.
Here the input text is represented at various levels that are spanning from coarser (e.g., sentences) to finer (e.g., phonemes) levels. 
Kenter et al.~\cite{kenter2019chive} proposed such kind of hierarchical approach to model prosodic features such as F0, energy, and duration, and these prosodic features along with linguistic features are utilized by a neural vocoder to render the final speech waveform.
However, one of the shortcomings of the prosody modelling studies based on latent representations is that they use reference mel-spectrograms to learn prosody embeddings during training, whereas during the inference time they either rely on textual based features to sample \cite{karlapati2020prosodic} or select \cite{tyagi2019dynamic} a prosody embedding from a set of pre-existing latent embeddings. This results in a mismatch between training and inference which could lead to an inappropriate prosody in the output speech signal.

Instead of modelling the prosodic latent space, few studies predict the conventional prosodic features (e.g., F0 and energy) in a multi-task manner and utilize those features to control the prosody of the synthesized speech~\cite{raitio2020controllable,ren2021fastspeech}.
The performance of these methods depends on the accuracy and robustness of prosodic feature extraction and modelling. 
However, F0 extraction and modelling is generally prone to a number of errors~\cite{drugman2018traditional}.

Complementing to the aforementioned studies, in this paper, we propose a Multi-Scale Spectrogram (MSS) modelling technique to capture short and long-range dependencies observed in the speech signal.
In MSS, the mel-spectrograms are predicted sequentially from a coarser scale capturing higher-level representation of speech to a finer scale capturing fine-grained prosodic details. Each subsequent finer scale is conditioned by the previous scale's predicted mel-spectrograms.
This allows the MSS modelling to produce prosody appropriate at different linguistic units such as sentence, word and phonemes, thereby improving the overall naturalness of the NTTS systems.

Similar to our proposed approach, Vasquez et al.~\cite{vasquez2019melnet} predict the mel-spectrograms in a multi-scale manner.
They initially predict lower resolution mel-spectrograms and progressively increase their resolution by 2 times at each scale irrespective of the semantic units in language or acoustic units in speech.
Contrary to that, we provide a generic multi-scale mechanism to represent mel-spectrograms and later develop two specific MSS systems where the scales correspond to linguistic units which are sentences, words, and phonemes. 
Moreover, each scale in MSS has its own objective to learn and all scales are learned together by multi-task learning~\cite{ruder2017overview}.

Another major difference between~\cite{vasquez2019melnet} and our approach is the use of explicit duration modelling instead of an attention mechanism to find alignment between text and speech.
Due to the stability issues of the attention mechanism of neural seq2seq models in the NTTS systems, the synthesized speech signals could have unpleasant artifacts such as mumbling, repetitions, or skipping \cite{he2019robust}. 
To mitigate the stability issues, non-attentive neural seq2seq models have recently become popular, where the attention mechanism is replaced by an explicit duration model~\cite{karlapati2020prosodic,zeng2020aligntts,ren2021fastspeech}.
Hence, the non-attentive neural seq2seq model based NTTS system is employed in this paper.

Our main contributions are as follows: i)~We propose a novel multi-scale mel-spectrogram modelling technique to improve the overall quality and naturalness of NTTS systems by appropriately capturing the coarse and fine-grained prosody;
ii)~We conduct and present an ablation study on two specific versions of MSS, called Sentence-level MSS and Word-level MSS. The Word-level MSS models word, phoneme, and frame-level spectrograms while Sentence-level MSS models sentence-level spectrogram additionally;
iii)~We evaluate Word-level MSS against a baseline system that is based on external duration model and show that it is significantly better than the baseline on two voices.

%% file: baseline.tex
\section{The Baseline}
\label{section:baseline}

\begin{figure}[htp]
    \centering
    \includegraphics[width=0.6\linewidth]{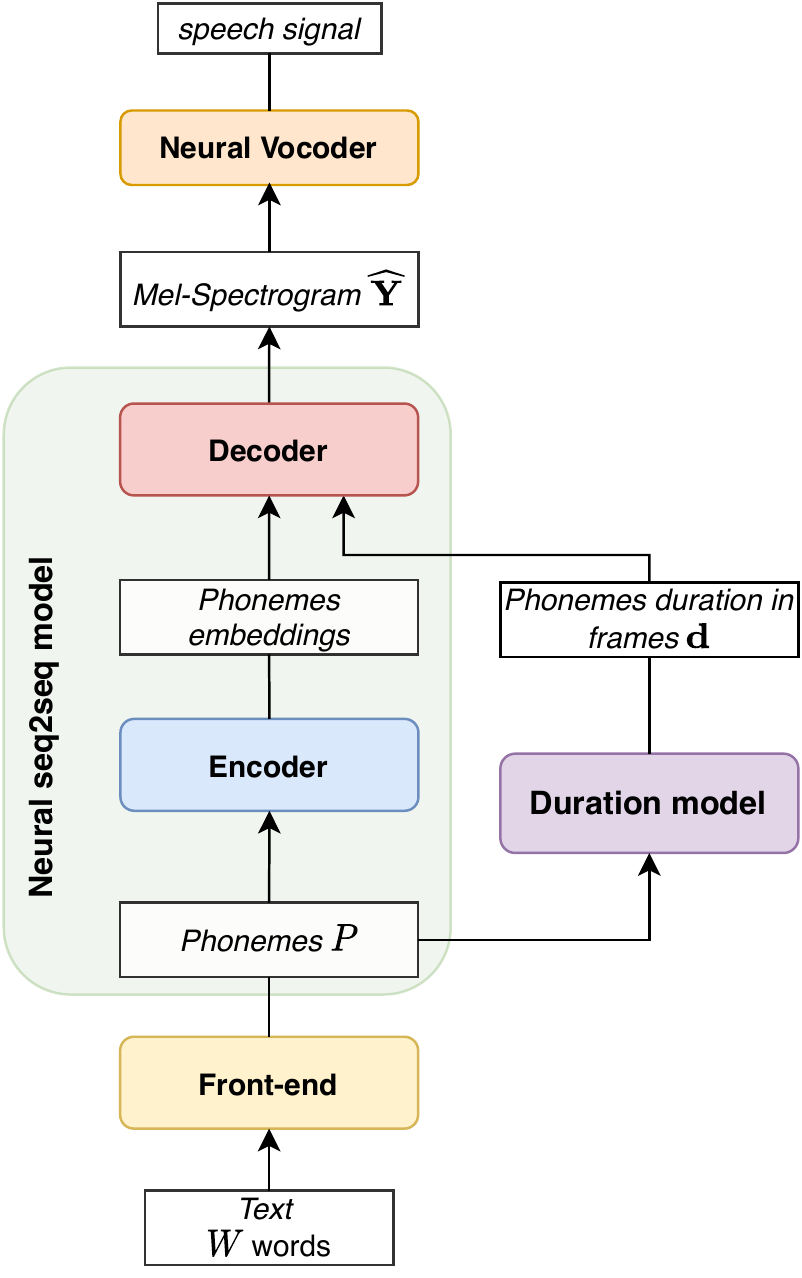}
    \caption{Block diagram showing the architecture of our baseline. The decoder module colored in red is substituted by a multi-scale decoder in the proposed MSS modelling technique.}
\label{fig:general_block_diagram}
\end{figure}

We use the same baseline system as in~\cite{karlapati2020prosodic}, which is a modified version of DurIAN~\cite{yu2019durian}. Figure~\ref{fig:general_block_diagram} illustrates the block diagram of our baseline system. It is composed of two major components: a seq2seq model and a duration model. 
First, the input text containing $W$ words $\vec{w}=[w_0, w_1, \dots, w_{W-1}]$ is passed through the front-end to extract $P$ phonemes $\vec{p}=[p_0, p_1, \dots, p_{P-1}]$ as an output. Next, the $P$ phonemes are passed through an encoder, which captures the relations between phonemes, and produces $P$ phoneme embeddings as an output. The encoder is composed of 1D convolutions followed by a bidirectional LSTM. Finally, the decoder takes these $P$ phoneme embeddings and $P$ phoneme durations in frames $\vec{d}=[d_0, d_1, \dots, d_{P-1}]$ where $\sum_{d \in \vec{d}}=T$ as an input, and predicts $T$ mel-spectrogram frames $\mathbf{Y}=[\vec{y}_0, \vec{y}_1, \dots, \vec{y}_{T-1}]$ auto-regressively as an output. 
There is no post-net after decoder as it resulted in instabilities when training with a reduction factor of 1.

During training, the phoneme durations are obtained from a forced-alignment algorithm, whereas during inference, the phoneme durations are predicted by a duration model trained separately. The duration model takes $P$ phonemes as an input and predicts $P$ durations.
Both the acoustic and the duration model are optimized using an L2 loss function.

%% file: mss.tex
\section{Multi-Scale Spectrogram (MSS) modelling} \label{sec:MSS}
\begin{figure}[tbp]
    \centering
    \includegraphics[width=0.95\linewidth]{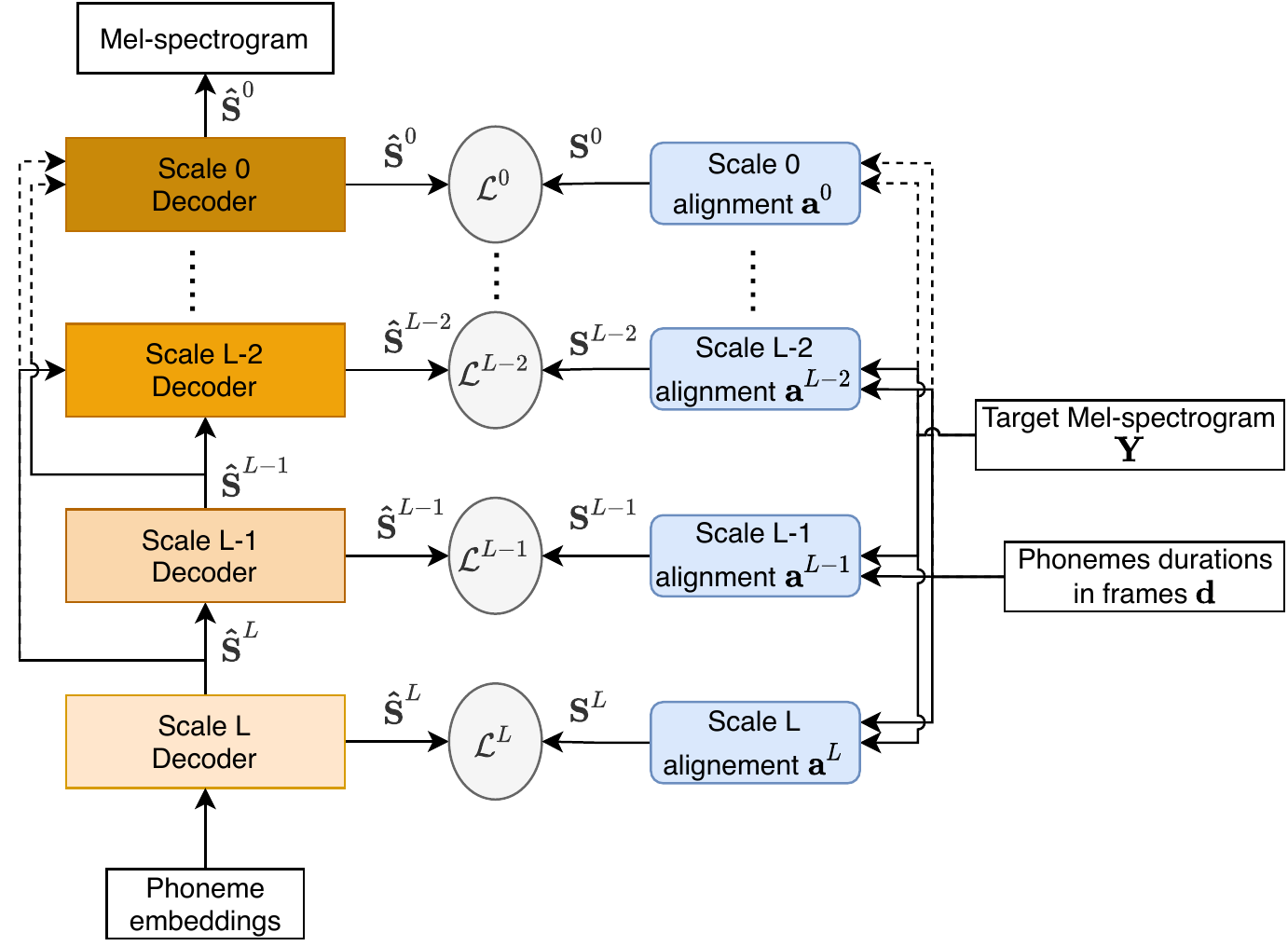}
    \caption{Block diagram of the generic MSS decoder with $L+1$ scales. The scale $0$ decoder depends upon the outputs of all the previous scale decoders.}
\label{fig:mss_decoder}
\end{figure}

This section introduces the proposed MSS modelling technique. As shown in Figure~\ref{fig:general_block_diagram}, the decoder of the baseline system predicts the mel-spectrogram frames directly from phoneme embeddings using phoneme durations. 
Contrary to this, the decoder based on MSS modelling technique predicts the mel-spectrogram frames after conditioning them on all the higher-level mel-spectrograms as illustrated in Figure~\ref{fig:mss_decoder}.
Specifically, the MSS modelling technique first predicts the mel-spectrogram vectors representing speech on a coarser scale which are later used for the prediction of mel-spectrogram vectors representing speech at a finer scale.
The coarser scale representation captures most of  the suprasegmental aspects of the prosody resulting in a more appropriate prosody for the given text. 
In principle, these scales can be defined in both time and frequency axes of the mel-spectrogram. 
However, in this paper, the scales are defined only along the time axis while keeping the number of mel-bins constant ($=80$) along the frequency axis. Extending the scales to the frequency axis is left as future work.

\subsection{Generic multi-scale representations} \label{subsec:mss_rep}

Before moving to modelling, we first discuss how to construct targets for learning a generic multi-scale model. Let us assume that there are in total $L+1$ scales in the MSS modelling. At each scale $l$ where $0 \leq l \leq L$, we compute a mel-spectrogram $\mathbf{S}^l=[\vec{s}^l_0, \vec{s}^l_1, \dots, \vec{s}^l_{N_l-1}]$ of dimension $N_l\times80$ from the ground-truth mel-spectrogram $\mathbf{Y}=[\vec{y}_0, \vec{y}_1, \dots, \vec{y}_{T-1}]$ of dimension $T\times80$.
Here, the $L^{\text{th}}$ scale (the highest level representation of speech) has the least number of mel-spectrogram vectors, and their number progressively increases on each subsequent scale $l<L$ until the last scale ${0}$ such that $N_L < N_{L-1} < \dots < N_0=T$. For example in Sentence-level MSS, $N_L=1$, which is the number of sentences in an utterance.

We compute the mel-spectrogram $\mathbf{S}^l$ at scales $l>0$ using the following equation:
\begin{equation*}
    \mathbf{S}^l=[\vec{s}^l_0, \vec{s}^l_1, \dots, \vec{s}^l_{N_l-1}],
\end{equation*}

where each 

\begin{equation}
\label{eq:mss_scale}
    \vec{s}^l_i= 
\begin{dcases}
    \frac{1}{a^l_i}\sum_{j=c^l_{i-1}}^{c^l_i}\vec{y}_j,& 1 \leq i < N_l,\\
    \frac{1}{a^l_i}\sum_{j=0}^{c^l_i}\vec{y}_j,& i=0,
\end{dcases}
\end{equation}

\begin{equation}
    c^l_k  = \sum_{i=0}^{k}a^l_i
\end{equation}

In Eq.~\ref{eq:mss_scale}, $\vec{a}^l = [a_0^l, a_1^l, \dots, a_{N_l-1}^l]$ is an alignment vector that denotes the alignments at scale $l>0$ where each $a^l_i$ represents the number of mel-spectrogram frames of $\mathbf{Y}$ corresponding to the spectrogram $\vec{s}_i^l$ at scale $l$. Thus the total sum of $\sum{\vec{a}^l} = T$. The alignment vectors $a^l_i$ can be computed based on the definition of each scale (cf. Section~\ref{subsection:architecture}).
The vector $\vec{c}^l = [c_0^l, c_1^l, \dots, c_{N_l-1}^l]$ is the cumulative sum of alignment vector $\vec{a}^l$, and each element $c^l_k$ corresponds to the starting frame for token $k-1$ and ending frame for token $k$.
So each target vector $\vec{s}^l_i$ at $l>0$ is computed by taking the mean of mel-spectrogram frames $\mathbf{Y}$ from index $c^l_{i-1}$ to $c^l_i$. The mel-spectrogram $\mathbf{S}^0$ at \nth{0} scale of MSS modelling technique is equal to the target mel-spectrogram $\mathbf{Y}$ i.e. $\mathbf{S}^0 = \mathbf{Y}$, thus $N_0 = T$.
This paper considers two specific cases of MSS modelling technique for validating our proposed approach: 1) \emph{Sentence-level MSS} and 2) \emph{Word-level MSS}, named after the coarser used linguistic unit.

\subsection{Word-level MSS and Sentence-level MSS}
\label{subsection:architecture}

The Word-level MSS has a total number of three scales ($L=2$) where the $\nth{1}$ and  $\nth{2}$ scales correspond to the linguistic unit of phonemes and words respectively.
The $\nth{0}$ scale corresponds to frame-level mel-spectrogram as discussed above. The number of mel-spectrogram vectors at each scale are given as such: $N_2 = \textrm{number of words}$ in a given utterance, $N_1 = \textrm{number of phonemes}$ present in an utterance, and $N_0=T$.
In Sentence-level MSS, $L=3$ and there is an additional \nth{3} scale that corresponds to sentences. The $N_3= \textrm{number of sentences}$ in a given utterance, which is equal to $1$ in our case. 
In both these specific systems, the alignment vectors $\vec{a}^l$ are obtained from the phoneme durations and relations between phoneme, word, and sentences, which are obtained by the front-end.
More specifically, in Sentence-level MSS, the alignment vector $\vec{a}^1$ is equal to phoneme durations $\vec{d}$ in frames. The alignment vector $\vec{a}^2$ is equal to the word durations in frames and $\vec{a}^3$ is equal to the sentence duration in frames.

The multi-scale representations that need to be modelled in Sentence-level MSS are $\mathbf{S} = [\mathbf{S}^0, \mathbf{S}^1, \mathbf{S}^2, \mathbf{S}^3]$.
To model $\nth{3}$ scale (sentence-level) mel-spectrogram $\mathbf{\hat{S}}^3$, we first project $P$ phoneme embeddings into a sentence-level vector by taking the last hidden state of the LSTM encoder.
Later, the sentence-level vector is passed through a sequence of 1D convolutions to obtain sentence-level mel-spectrogram $\mathbf{\hat{S}}^3$.
The loss function at the $\nth{3}$ scale (sentence-level) is defined as:
\begin{equation}
    \mathcal{L}^3 = \norm{\mathbf{\hat{S}}^3 - \mathbf{S}^3}_2
\end{equation}
The $\mathbf{\hat{S}}^3$ mel-spectrogram is assumed to capture the sentence-level acoustic properties such as speaker-identity, recording environment, or speaking style of the sentence.

Similarly, to model the $\nth{2}$ scale (word-level) mel-spectrogram $\mathbf{\hat{S}}^2$, we first project phonemes of each word into a word-level vector. 
Later, the word-level vectors are concatenated with the upsampled sentence-level mel-spectrogram $\mathbf{\hat{S}}^3_\uparrow$. $\mathbf{\hat{S}}^3_\uparrow$ is computed by upsampling the predicted sentence-level mel-spectrogram $\mathbf{S}^3$ to have the same dimension as $\mathbf{S}^2$ using an alignment defined between $\nth{2}$ and $\nth{3}$ scale, i.e. between words and the sentence they belong to.
The loss function on \nth{2} scale is defined as:
\begin{equation}
    \mathcal{L}^2 = \norm{\mathbf{\hat{S}}^2 - \mathbf{S}^2}_2
\end{equation}

$\mathbf{\hat{S}}^2$ is assumed to capture the word level acoustic properties such as word prominence, rise and fall of pitch and energy at word level.

We follow the same procedure to predict the phoneme-level mel-spectrogram $\mathbf{\hat{S}}^1$ and final frame-level mel-spectrogram $\mathbf{\hat{S}}^0$ or $\mathbf{\hat{Y}}$. In each of the scales $l$, the predicted mel-spectrograms are also conditioned on all the coarser scale predicted mel-spectrograms. The loss function at each scale $l$ is defined as $\mathcal{L}^l = \norm{\mathbf{\hat{S}}^l - \mathbf{S}^l}_2$. The sentence-level MSS is trained by minimizing the loss at all scales. So the training loss for the whole system is defined as:

\begin{equation}
    \label{eq:mss_loss}
        \mathcal{L} = \mathcal{L}^0 + \mathcal{L}^1 + \mathcal{L}^2 + \mathcal{L}^3 ,
    \end{equation}

which can be interpreted as maximizing the likelihood of mel-spectrograms at all scales:

\begin{equation}
\begin{split}
    p(\mathbf{\hat{S}}^3,\mathbf{\hat{S}}^2,\mathbf{\hat{S}}^1,\mathbf{\hat{S}}^0)=&
    p(\mathbf{\hat{S}}^3).p(\mathbf{\hat{S}}^2|\mathbf{\hat{S}}^3).p(\mathbf{\hat{S}}^1|\mathbf{\hat{S}}^3,\mathbf{\hat{S}}^2)\\&.p(\mathbf{\hat{S}}^0|\mathbf{\hat{S}}^3,\mathbf{\hat{S}}^2,\mathbf{\hat{S}}^1)
\end{split}
\end{equation}

%% file: experiments.tex
\section{Experiments}
\label{section:experiments}

    \subsection{Data}
    \label{subsection:data}
    
    The experiments were carried out on an internal voice dataset that was recorded by two native US-English female voice talents. We refer to them as speaker-A and speaker-B.
    The training and test sets for speaker-A are 33 and 5.5 hours respectively, while they are 32 and 3 hours for speaker-B respectively. The test set is reserved for testing in this and future research studies. The sampling rate of the recorded audio is 24kHz. We extracted 80 band mel-spectrograms with a frame shift of 12.5ms.

    \subsection{Training and inference}

    As mentioned in Section \ref{subsection:architecture}, we have developed two systems based on MSS modelling technique: Sentence-level MSS and Word-level MSS. In both systems:~1)~we train the seq2seq model according to the loss defined in Equation~\ref{eq:mss_loss} where oracle alignments~$\vec{a}^l$ are provided to the model at each scale $l$ ; 2)~we train the duration model using L2 loss as shown in Section~\ref{section:baseline}.
    We optimize the loss in both the models using Adam optimizer~\cite{kingma2014adam} with different learning rates. We use a learning rate of $10^{-3}$ and $10^{-4}$ for the acoustic and the duration model respectively.

    During inference, we follow these 2 steps in the following order: 
    I)~we predict durations $\vec{\hat{d}}$ from the duration model trained in step 2; II)~we generate mel-spectrograms $\mathbf{\hat{S}}^{0}$ from the seq2seq model trained in step 1 using the predicted durations $\vec{\hat{d}}$ from step I. We use Wavenet vocoder~\cite{oord2016wavenet} to synthesize speech from $\mathbf{\hat{S}}^{0}$.

    \subsection{Evaluations}

    \subsubsection{Qualitative evaluation of predicted and target spectrograms on different scales}

\begin{figure}[tp]
        \centering
        \begin{subfigure}{.49\linewidth}
            \includegraphics[width=\linewidth]{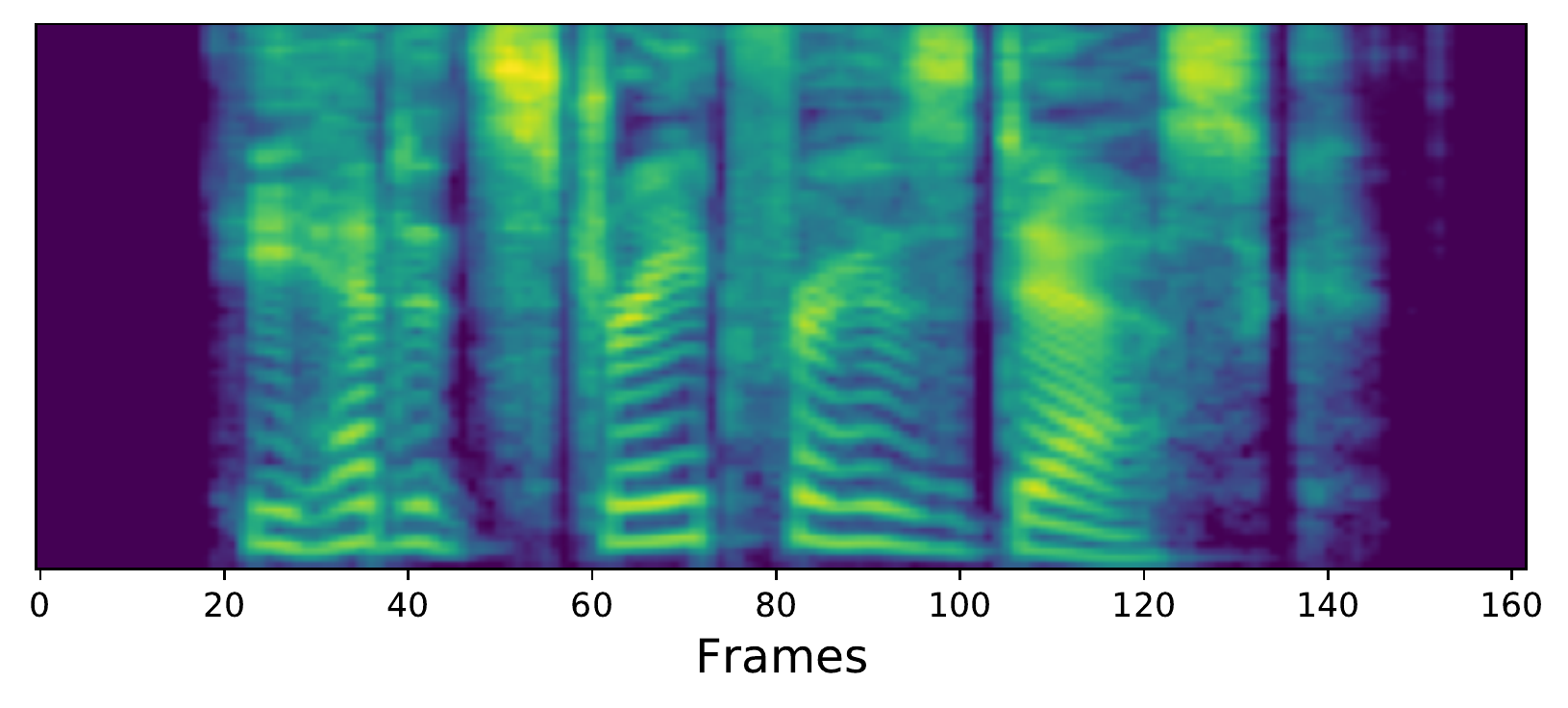}
        \end{subfigure}
        \begin{subfigure}{.49\linewidth}
            \includegraphics[width=\linewidth]{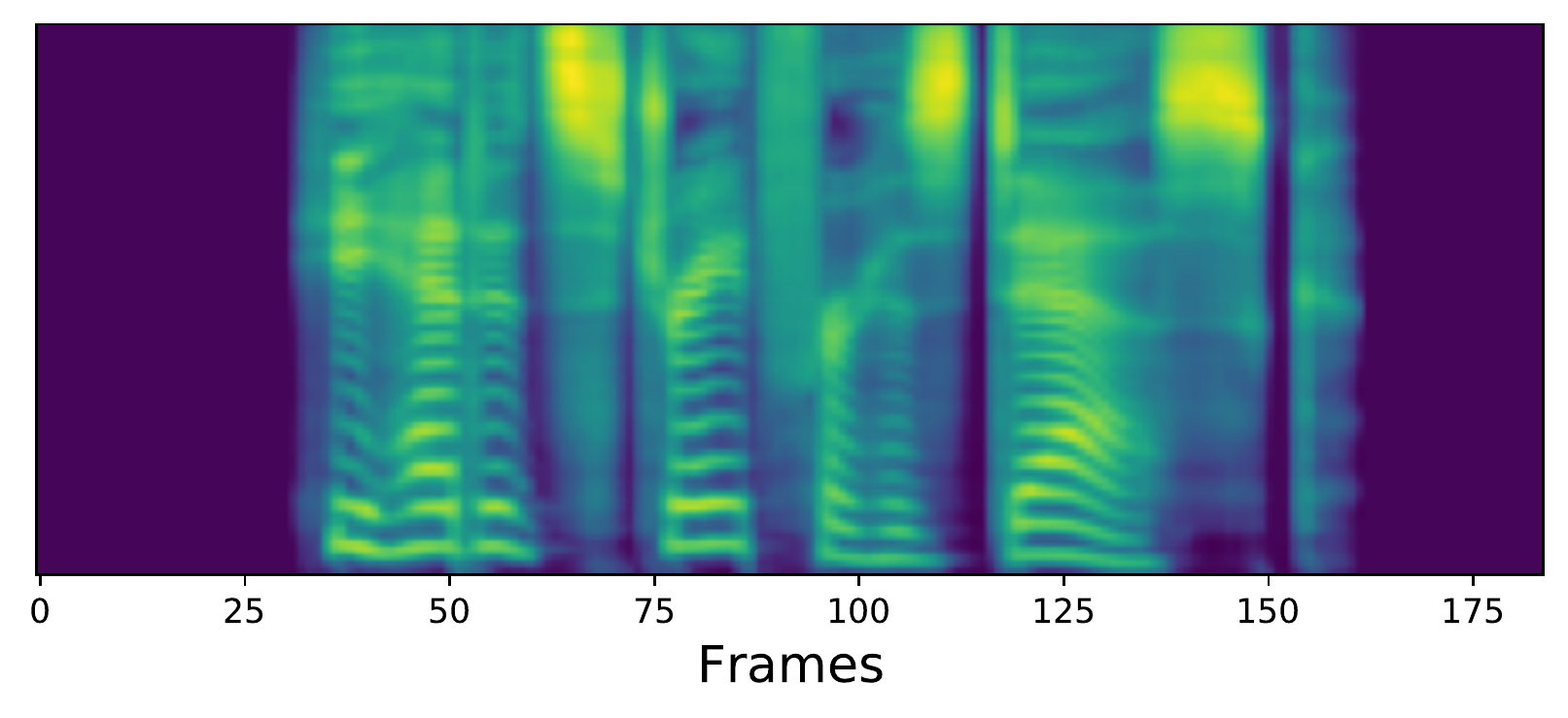}
        \end{subfigure}
        \begin{subfigure}{.49\linewidth}
            \includegraphics[width=\linewidth]{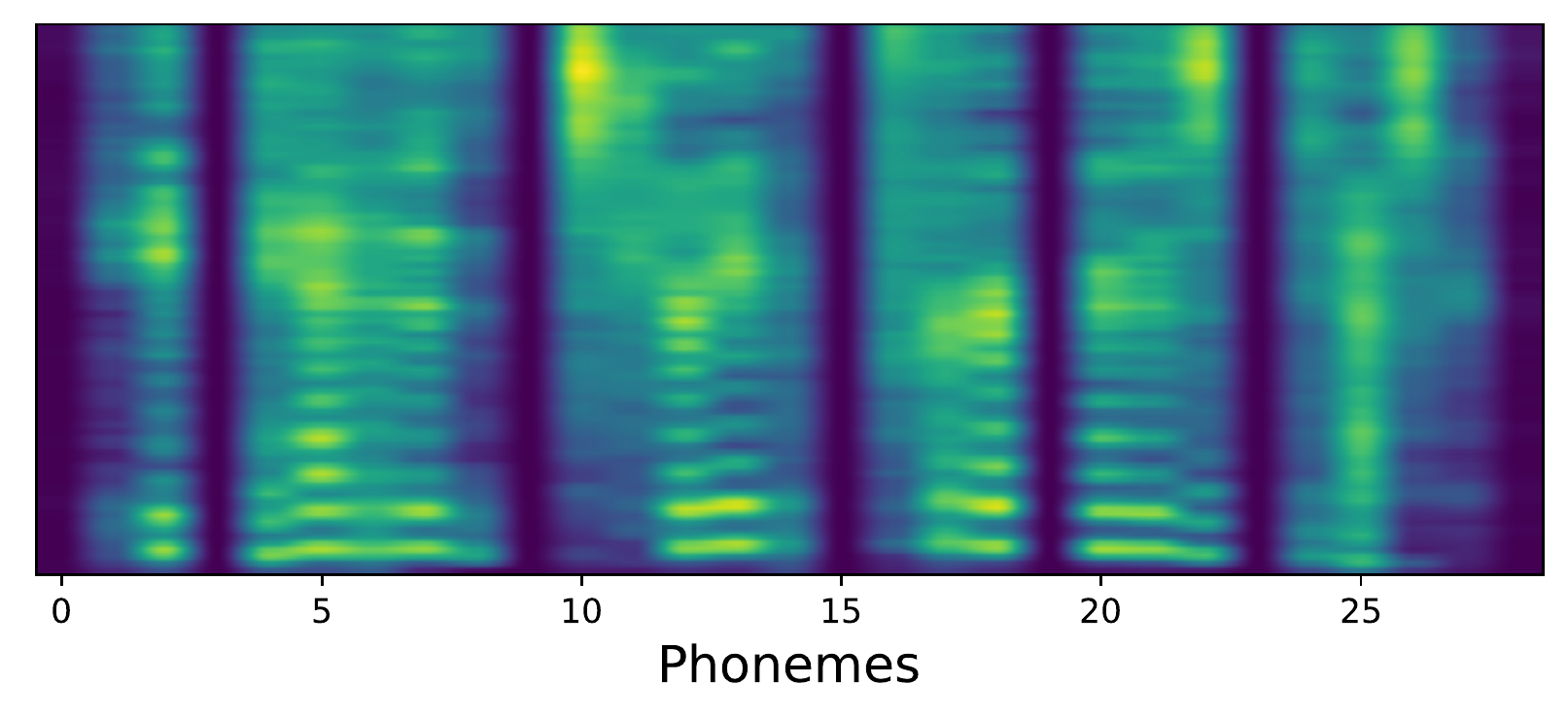}
        \end{subfigure}
        \begin{subfigure}{.49\linewidth}
            \includegraphics[width=\linewidth]{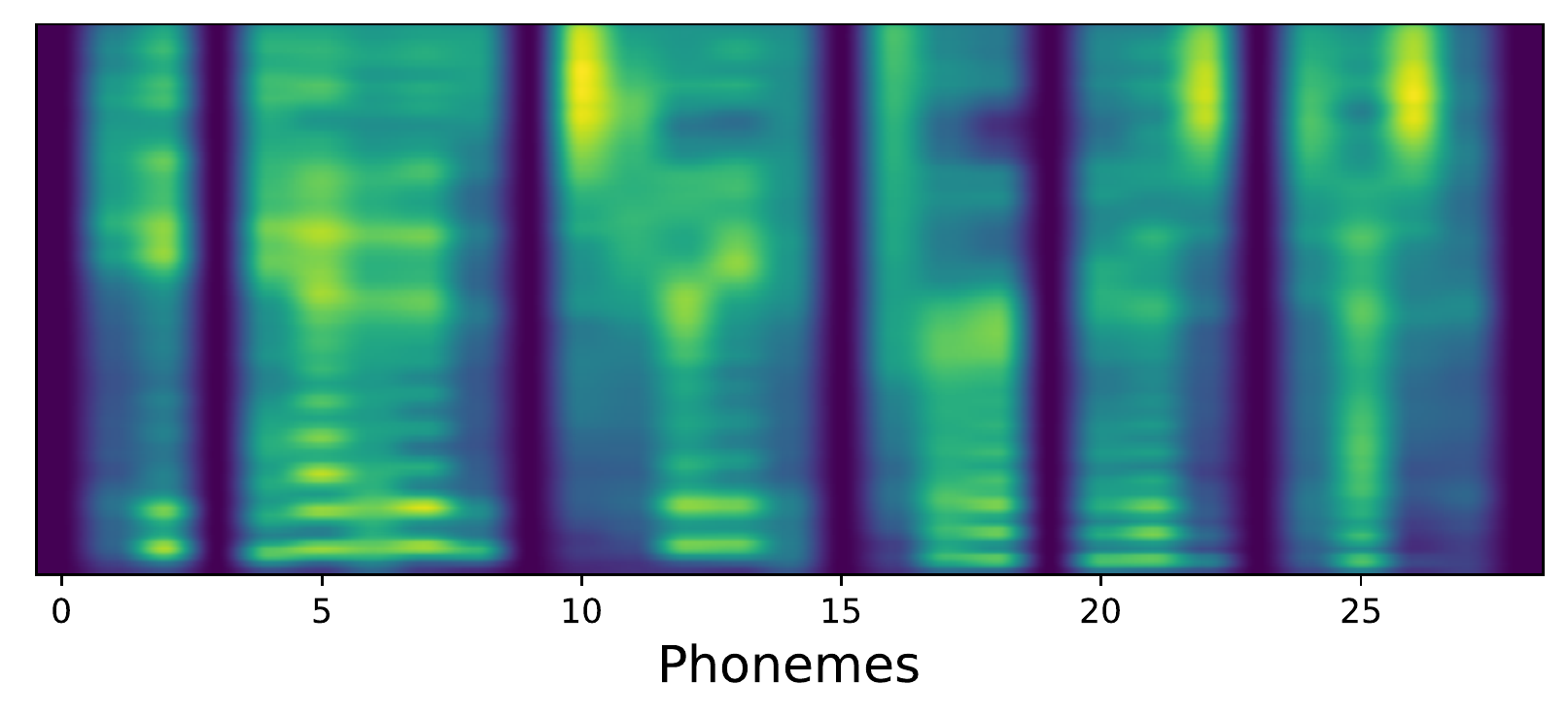}
        \end{subfigure}
        \begin{subfigure}{.49\linewidth}
            \includegraphics[width=\linewidth]{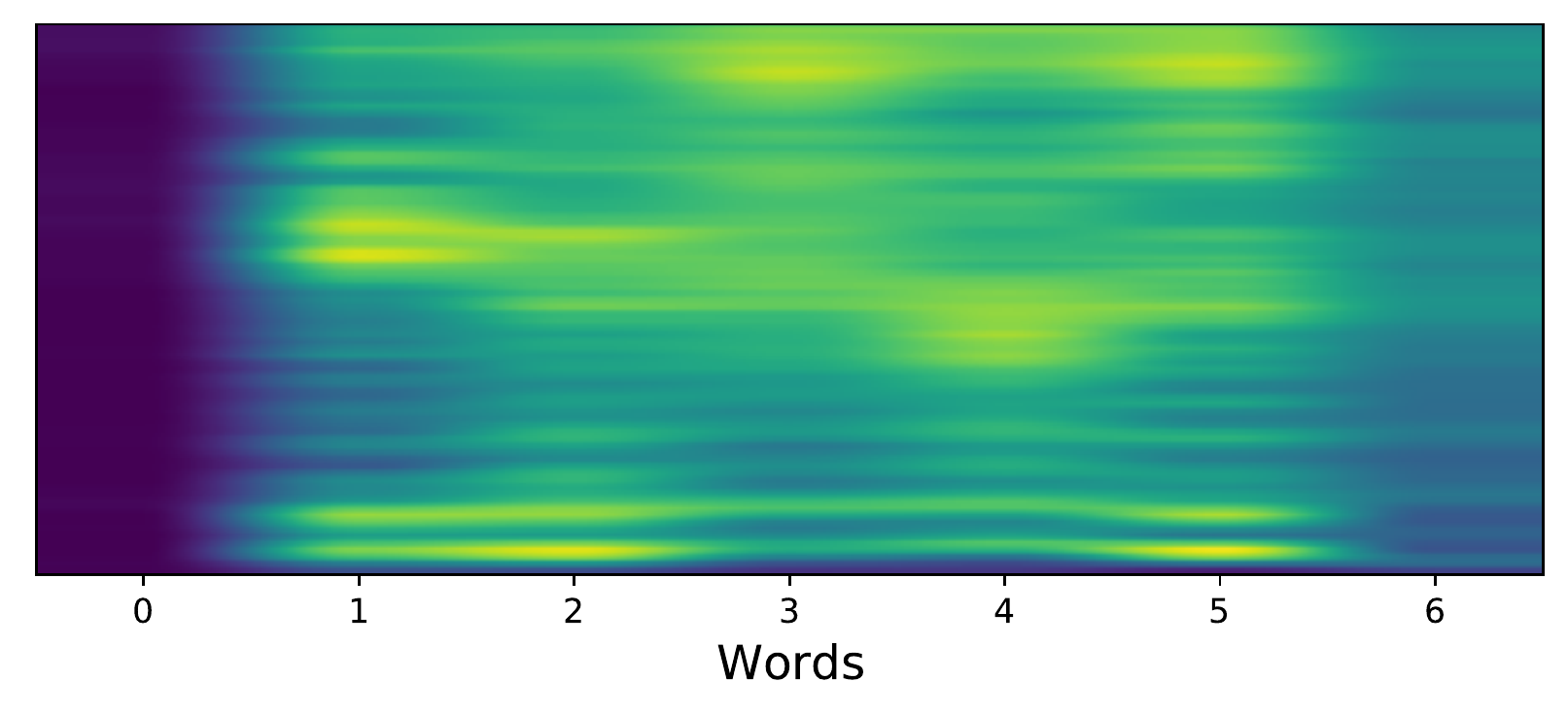}
            \caption{Oracle spectrogram}
        \end{subfigure}
        \begin{subfigure}{.49\linewidth}
            \includegraphics[width=\linewidth]{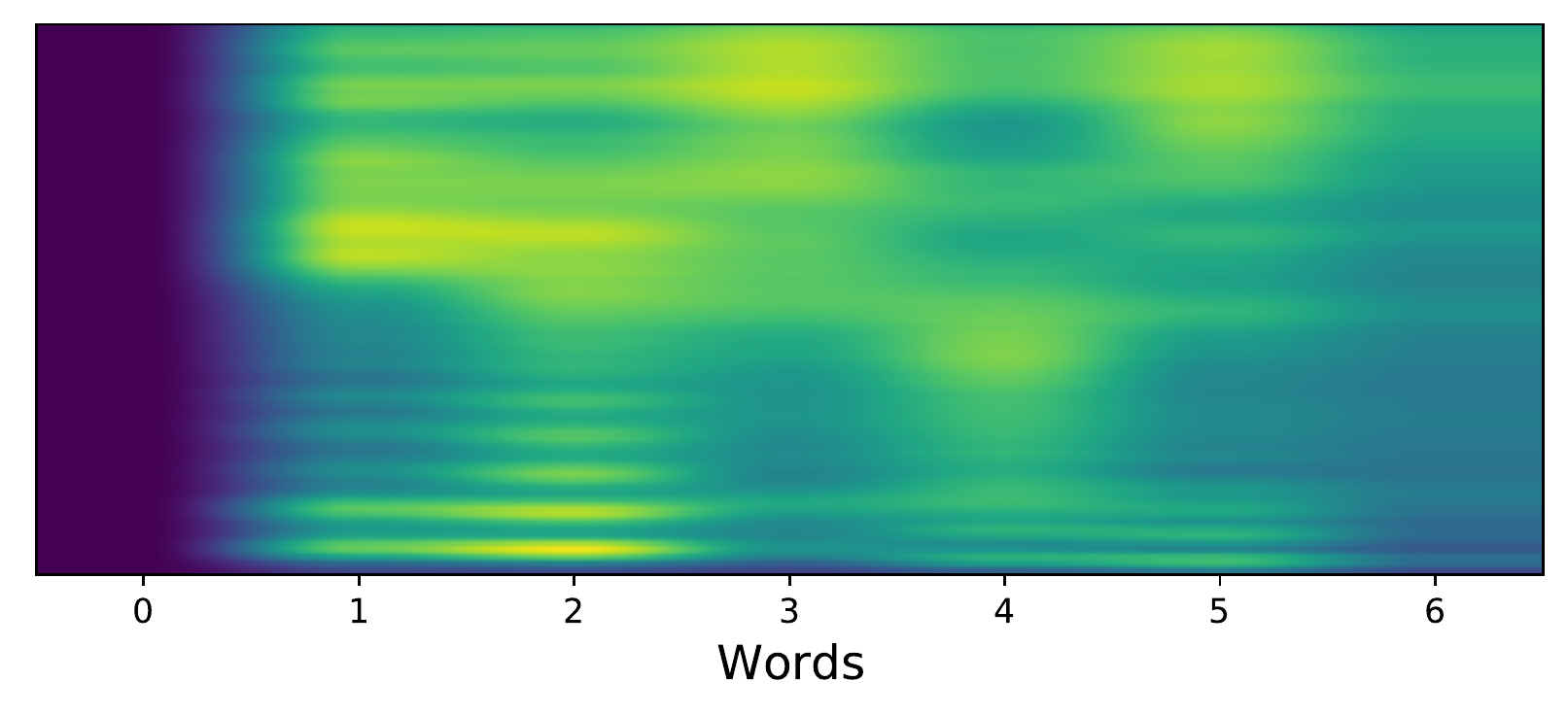}
            \caption{Predicted spectrogram}
        \end{subfigure}
        \caption{Visualisation of mel-spectrograms at different scales in Word-level MSS given the text: ``He headed straight for his desk.''. Left panel: oracle spectrograms, right panel: predicted spectrograms. The mel-spectrograms in bottom, middle, and top row correspond to \nth{2} (word), \nth{1} (phoneme), and \nth{0} (frame) scale respectively.}
        \label{fig:visualisation_mel_specs}
    \end{figure}

    Figure~\ref{fig:visualisation_mel_specs} shows the target (left column) and predicted (right column) mel-spectrograms at different scales $l$ in the Word-level MSS for speaker-B as an example. The bottom row shows the $\nth{2}$ scale (word-level) mel-spectrogram $\mathbf{S}^2$. There are 7 mel-spectrogram vectors $[\vec{s}_0^2, \vec{s}_1^2, \dots, \vec{s}_6^2]$ which represent coarse-grained prosodic features such as prominence at word-level. We can observe the harmonics and energy, and how they vary after each word. The middle row shows the phoneme-level mel-spectrogram $\mathbf{S}^1$. On this scale, there are 29 mel-spectrogram vectors $[\vec{s}_0^1, \vec{s}_1^1, \dots, \vec{s}_{28}^1]$ which represent fine-grained prosodic features at phoneme-level. At this scale, we can see how the prosody varies around phonemes and can identify the stress on phonemes based on their acoustic representations. The top row shows frame-level mel-spectrogram $\mathbf{S}^0$, which is equal to the target spectrogram $\mathbf{Y}$. 
    When comparing the target (left column) to the predicted (right column) mel-spectrograms, the Word-level MSS is able to capture the high-level prosodic features at $\nth{1}$ and $\nth{2}$ scale, albeit with a smoother representation due to the nature of the L2 loss used in eq.~\ref{eq:mss_loss}.

    \subsubsection{Ablation study}
    \label{subsection:ablation_scales}

\begin{figure}[tp]
        \centering
        \includegraphics[width=0.98\linewidth]{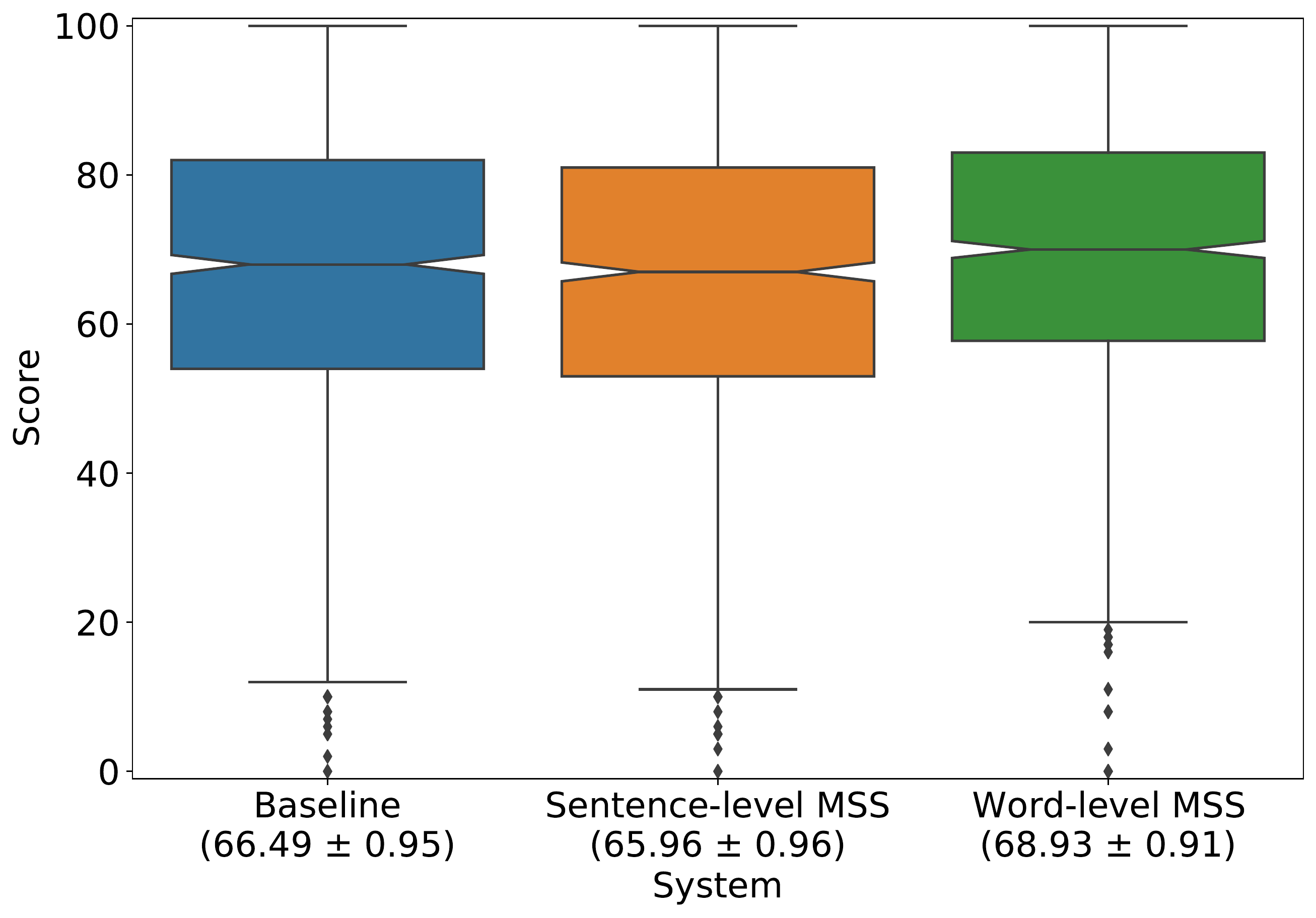}
        \caption{Results of the MUSHRA evaluation of ablation study on speaker-A voice. Mean rating and 95\% confidence intervals are reported below system names.}
    \label{fig:mushra_mss_ablations}
    \end{figure}

    A MUSHRA~\cite{series2014method} evaluation was conducted on speaker-A to evaluate how the number/definition of scales affects the performance of MSS modelling. For the MUSHRA evaluation, we selected the following three systems: the baseline from Section~\ref{section:baseline}, Word-level MSS, and Sentence-level MSS. A total of 50 utterances were selected randomly from the test set, and the duration of each utterance was approximately 15 seconds. Each utterance was rated by 24 native US-English professional listeners.
    Figure~\ref{fig:mushra_mss_ablations} presents the results of the MUSHRA evaluation.  
    We have used a pairwise two-sided Wilcoxon signed-rank test corrected for multiple comparisons to measure statistical significance between the systems. The Word-level MSS system performed statistically significantly better than both the other systems ($p$-value $< 10^{-6}$) while there was no statistically significant difference between the baseline and sentence-level MSS systems ($p$-value $=0.27$). We believe that during training, the Sentence-level MSS system is overfitting on the sentence-level spectrogram $\mathbf{\hat{S}}^3$ prediction. Due to this, it fails to capture coarse-grained prosodic features observed in the target $\mathbf{S}^3$, thus adversely affecting the prediction of spectrograms in lower scales $l<L$ which are conditioned on $\mathbf{\hat{S}}^3$. Moreover, an utterance in our data corresponds to a sentence which makes the prediction of $\mathbf{\hat{S}^3}$ even more difficult because it does not have the surrounding sentences as an input to the system unlike scales $l<L$. These reasons could suggest why there is no improvement in the Sentence-level MSS system compared to the baseline.

    \subsubsection{Preference tests}
\begin{figure}[tp]
        \centering
        \begin{subfigure}{\linewidth}
            \includegraphics[width=\linewidth]{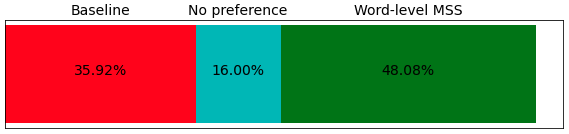}
            \caption{speaker-A}
            \label{fig:pref_word_mss_spkA}
        \end{subfigure}

        \begin{subfigure}{\linewidth}
            \includegraphics[width=\linewidth]{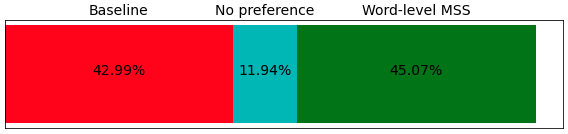}
            \caption{speaker-B}
            \label{fig:pref_word_mss_spkB}
        \end{subfigure}
        \caption{The results of preference tests between the Word-level MSS and the baseline system on two voices.}
    \end{figure}

    As the Word-level MSS system showed a significant improvement in the ablation study, we have selected it for further comparisons with the baseline system.
    A preference test was conducted on speaker-A and speaker-B voices. 
    For speaker-A, 50 utterances were selected randomly from the test set and each utterance had a duration of approximately 15 seconds.
    Similar to the MUSHRA evaluation, we used a third-party vendor to complete the test, and a total 24 listeners participated. 
    The results are shown in Figure~\ref{fig:pref_word_mss_spkA}. We use a binomial significance test to measure  the statistical significance. The Word-level MSS is found to be statistically significantly better than the baseline ($p$-value $<10^{-4}$)

    For speaker-B, 100 utterances were selected randomly from the test set and the duration of each utterance was approximately 15 seconds. A total of 48 subjects participated in the preference test. However, this evaluation was conducted via Clickworker platform - a crowdsourced evaluation, unlike earlier evaluations. The results of the preference test are shown in Figure~\ref{fig:pref_word_mss_spkB}. There was not a statistically significant preference for the Word-level MSS system ($p$-value=$0.068$). 
    However, upon removing listeners that had low reliability because they did not listen completely to both samples, we found that the difference becomes significant, i.e. Word-level MSS is preferred statistically significantly ($p$-value=$0.007$).
    The results of both MUSHRA and preference tests suggest that the Word-level MSS system is able to produce more natural speech than the baseline system.
    We observe that the Word-level MSS has more contextually appropriate emphasis on the words and generally better intonation without impacting the segmental quality. Although the Word-level MSS system is preferred over the baseline system, we do note that on certain utterances it is unable to produce the right kind of intonation, specifically when a question does not start with an interrogative word.

%% file: conclusions.tex
\section{Conclusion}

    In this paper, we presented a novel method for multi-scale modelling of mel-spectrograms to improve the quality of NTTS systems. 
    We presented a generic MSS modelling approach and later provided details for its two specific versions called Sentence-level MSS and Word-level MSS where the scales correspond to the linguistic units. The ablation study showed that the Word-level MSS system performed statistically significantly better than Sentence-level MSS. In the preference evaluations on 2 voices, the Word-level MSS system showed statistically significantly better results than the baseline system. In the future, we want to introduce scales in MSS along the frequency axis as well which could result in an even improved segmental quality. We also want to extend the sentence-level MSS to broader linguistic units for a better modelling of the coarse-grained prosody of speech. Furthermore, we want to introduce another scale that corresponds to syllables as they are strongly linked to prosodic events like stress and intonation~\cite{ito2018syllable}.